\documentclass[useAMS,usenatbib]{mn2e}
\usepackage{placeins}
\usepackage{epsf}
\usepackage{amssymb}
\usepackage{amsmath}
\usepackage[usenames]{color}
\usepackage{graphicx}

\def\deg{$^{\circ}$}
\def\co56{$^{56}{\rm Co}$}
\def\ni56{$^{56}{\rm Ni}$}
\def\fe56{$^{56}{\rm Fe}$}

\newcommand{\be}{\begin{equation}}
\newcommand{\ee}{\end{equation}}

\def\***#1{\textbf{\textsf{*** #1 ***}}}

\newcommand {\INTEGRAL}{{\it INTEGRAL }}
\newcommand {\ASTROGAM}{{\it ASTROGAM }}

\title[Polarization of gamma-rays from SN Ia]{Polarization of MeV gamma-rays and 511 keV line shape as probes of SNIa asymmetry and magnetic field}
\author[Churazov et al.]{E.~Churazov$^{1,2}$ \& I.~Khabibullin$^{1,2}$
 \newauthor \\
$^1$ Max-Planck-Institut f\"ur Astrophysik, Karl-Schwarzschild-Strasse 1, 85741
Garching, Germany\\
$^2$ Space Research Institute (IKI), Profsoyuznaya 84/32, Moscow 117997, 
Russia\\
}

\begin{document}

\pagerange{\pageref{firstpage}--\pageref{lastpage}}

\maketitle

\label{firstpage}
\begin{abstract}
We discuss  gamma-ray signatures associated with an asymmetric explosion and transport of positrons in SN Ia ejecta. In particular,  Compton scattering of gamma-ray line photons can induce polarization in the continuum, which would be a direct probe of the asymmetries in the distribution of radioactive isotopes and/or of the scattering medium. Even more interesting would be a comparison of the shapes of $\gamma$-ray lines and that of the electron-positron annihilation line at 511~keV. The shapes of $\gamma$-ray lines associated with the decay of  \co56 (e.g., lines at 847 and 1238~keV) directly reflect the velocity distribution of \co56. On the other hand, the 511~keV line arises from the annihilation of positions, which are  also produced by the \co56 decay but can propagate through the ejecta before they slow down and  annihilate. Therefore, the shape of the annihilation line might differ from other gamma-ray lines, providing constraints on the efficiency of positrons propagation through the ejecta and, as consequence, on the  topology of magnetic fields in the ejecta and on the fraction of positrons that escape to the interstellar medium. We illustrate the above effects with two models aimed at capturing the main predicted signatures.  
\end{abstract}


%

\sloppypar

\section{Introduction}

Besides serving as an important tool for cosmology, thermonuclear (Type Ia) supernovae (SN) are extremely interesting astrophysical objects by themselves \citep[e.g.][]{2000ARA&A..38..191H,2005AstL...31..528I}. While the thermonuclear nature of SNIa's \citep{1960ApJ...132..565H} is beyond doubts, there are ongoing debates on their progenitors \citep[see, e.g.,][for a review]{2012NewAR..56..122W}, triggering mechanisms of the explosion and details of the combustion and burning dynamics. It is plausible that there are several different channels leading to SNIa-type events, further complicating these debates.  Thousands of Type Ia supernovae have already been detected in and monitored in the optical band. The optical emission arises from re-processing of gamma-ray photons and positrons, copious amounts of which are produced by the radioactive decay of the nuclear burning end-products, namely  \ni56 and then \co56. Modeling of SNIa optical spectra is, however, complicated, even if the geometry of the problem is specified, since one has to calculate detailed ionization and thermal balance of the ejecta and take into account multitude of lines of weakly ionized spices like iron, silicon, etc.  

From the radiative transfer point of view, the gamma-ray band is much more straightforward to model \citep[e.g.][]{1969ApJ...155...75C,2014ApJ...786..141T}. Indeed, it involves only photoelectric absorption and Compton scattering (ignoring the contribution
of positrons for a moment). For the ``late'' phase of the SN expansion 
(50-100 days after the explosion), when optical depth of the ejecta for gamma-ray photons drops below
unity, even a single scattering approximation provides reasonably accurate results.

Only one Type Ia supernova (SN2014J) has been detected and extensively studied in the gamma-ray band so far. The fluxes and widths of gamma-ray lines provided the most direct confirmation of the thermonuclear nature of SN2014J. The derived parameters values turned out to be close to the canonical values: the total mass $\sim$ Chandrasekhar mass and the radioactive \ni56 mass $\sim$ 0.6 $M_\odot$ \citep[e.g.][]{2014Natur.512..406C,2015ApJ...812...62C}, consistent with the basic spherically symmetric models  \citep[see, however, ][for possible peculiarities]{2014Sci...345.1162D,2016A&A...588A..67I}.

Here we discuss two effects that can be used to probe the asymmetry of the ejecta and, indirectly, the topology of the magnetic field inside it. The first one involves energy-dependent polarization of the gamma-ray emission arising due to asymmetries in the ejecta. The second one links the shape of the electron-positron annihilation line at 511 keV with positrons' ability to migrate through the ejecta and eventually escape.  

The structure of the paper is as follows. In Section~\ref{sec:pol} we consider polarization induced by an aspherical SNIa ejecta in the scattered continuum. In Section~\ref{sec:pos} we consider the changes in the 511 keV line caused by propagation of positrons through the ejecta. Section~\ref{sec:conclusions} summarizes our findings. 

\section{Gamma-ray polarization}
\label{sec:pol}
Polarization of optical emission is a standard tool in supernova studies \citep[e.g.][]{1982ApJ...263..902S,1991A&A...246..481H,1992SvAL...18..168C,2003ApJ...593..788K,2008ARA&A..46..433W,2016ApJ...831...79I}.
It arises due to resonant or Thomson scattering of photons propagating through an aspherical atmosphere \citep[e.g.][]{1977A&A....57..141B}. In gamma-rays, the asphericity of the ejecta should also lead to the polarization, which is going to be energy dependent due to the recoil effect.
Primary gamma-rays are produced as narrow lines, which are then broadened by the velocities of homologously expanding ejecta. Between 50 and few hundred days after the explosion, the most bright are the lines at 847 and 1238~keV, produced by \co56 decay into stable \fe56. If one ignores the effects associated with positrons, which are produced in $\sim19$\% of decays, the unscattered gamma-spectrum should consist of the gamma-ray lines, whose shapes reflect the velocity distribution of \co56. In principle, strong asymmetry in the \co56 distribution can already reveal itself in the line profiles. Scattering of line photons generates low-energy continuum/wings of the lines due to recoil effect. In this scattered continuum, the asymmetry of either the distribution of the decaying \co56 or of the scattering medium should lead to polarization. We illustrate the above discussion with a set of models, which are aimed to single out the effects of
induced polarization.

\subsection{Homogeneous ellipsoid with a point source at the center}
\label{sec:el}
Consider a homogeneous ellipsoid of rotation (with semi-axes $b_x=1$ and $b_y=b_z=b$) with an isotropic source of monochromatic gamma-ray photons of energy $E_0$ at its center. The line of sight is along the $z$ axis of the ellipsoid, while $x$ and $y$ are in the sky plane. We consider only pure Compton scattering of gamma-ray photons on free electrons at rest and ignore the broadening caused by motions of the ejecta. Time delays due to the propagation of photons from different parts of the ejecta are also neglected. 
We further assume that the ellipsoid is optically thin for gamma-rays, which for realistic SNIa models corresponds to late phases of the expansion, say, after 50-100 days. If so, the probability of scattering is proportional to the length $a(\theta,\phi)$ of the vector from the center of the ellipsoid to its surface, where  $\theta$ is the angle with respect to the $z$-axis, i.e., the scattering angle, and $\phi$ is the angle of rotation around the line of sight. For the ellipsoid
\be
a(\theta,\phi)=\frac{b}{\left[(b^2-1)\cos^2 \phi \sin^2\theta +1 \right]^{1/2}}.
\ee 
Given that the energy of the photon after the scattering 
\begin{eqnarray}
E_{obs}&=&\frac{E_0}{\left [1+\frac{E_0}{m_ec^2}\left (1-\cos \theta \right) \right ]} 
\label{eq:eobs}
\end{eqnarray}
depends only on $\theta$ and not on $\phi$, the observed spectrum of scattered emission (in a single-scattering approximation) can be written as a function of $\theta$ or $E_{obs}$ as $\propto\displaystyle \int \frac{d\sigma}{d\Omega}_{KN} a(\theta,\phi)d\phi$, where $\displaystyle \frac{d\sigma}{d\Omega}_{KN}=\frac{1}{4}r_0^2 X^2 \left [X+X^{-1}-2+4 \cos^2\Theta \right]$
is the Klein-Nishina cross section for polarized photons; $\displaystyle X=E_{obs}/E_0$ and $\Theta$ is the angle between the polarization vectors of the photon before and after the scattering.  Given the symmetry of the problem only two Stokes parameters  $I$ and $Q$ are needed to completely specify the polarization of the final state. We define the degree of polarization as $P=Q/I$ [note, that in the SN~Ia spectropolarimetry literature the convention is to use another definition,   $\displaystyle P=(Q^2+U^2)^{1/2}/I$]. 
Here $Q$ is defined with respect to the $x$ and $y$ axis in such a way, that when the photon is polarised along the $y$ axis, $Q=I$ and, therefore, $P=1$ (correspondingly, if it is polarised along the $x$ axis, $P=-1$).

\begin{figure}
\begin{minipage}{0.49\textwidth}
\includegraphics[trim= 1mm 5cm 20mm 2cm,
  width=1\textwidth,clip=t,angle=0.,scale=0.98]{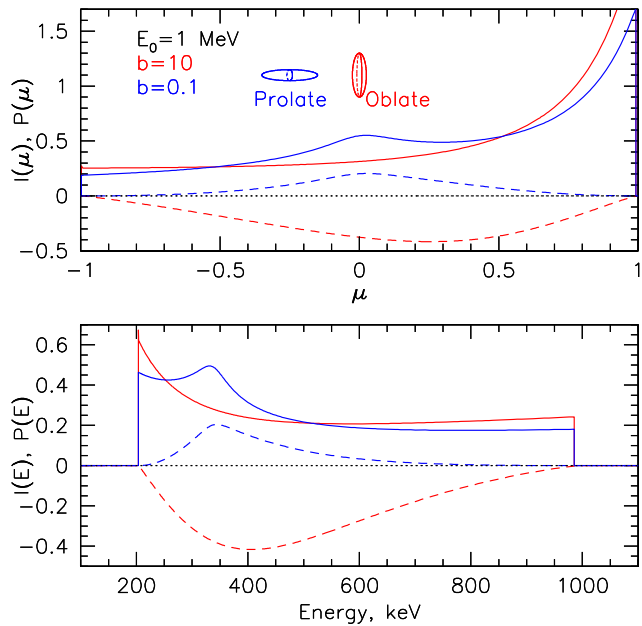}
\end{minipage}
\caption{ Intensity $I$ (solid lines) and the degree of polarization $P$ (dashed lines) of gamma-ray emission scattered on a homogeneous ellipsoid of rotation with a central point source of monochromatic photons with initial energy $E_0=1$\,MeV. The top panel shows $I(\mu)$ and $P(\mu)$, with $\mu$ being cosine of the scattering angle, while the bottom panel shows the same quantities as functions of photon energy, which can be measured by an observer.  Extreme values of $b=10$  (red lines) and 0.1 (blue lines)  are used for illustration (see inset schematics of these cases). For the prolate ellipsoid, scattering angles $\sim$90\deg dominate, leading to a peak at $\mu\sim0$ and at $E\sim 340$\,keV. These features are also reflected in the polarization degree.
\label{fig:pol_spec}
}
\end{figure}

\begin{figure}
\begin{minipage}{0.49\textwidth}
\includegraphics[trim= 1mm 5cm 20mm 2cm,
  width=1\textwidth,clip=t,angle=0.,scale=0.98]{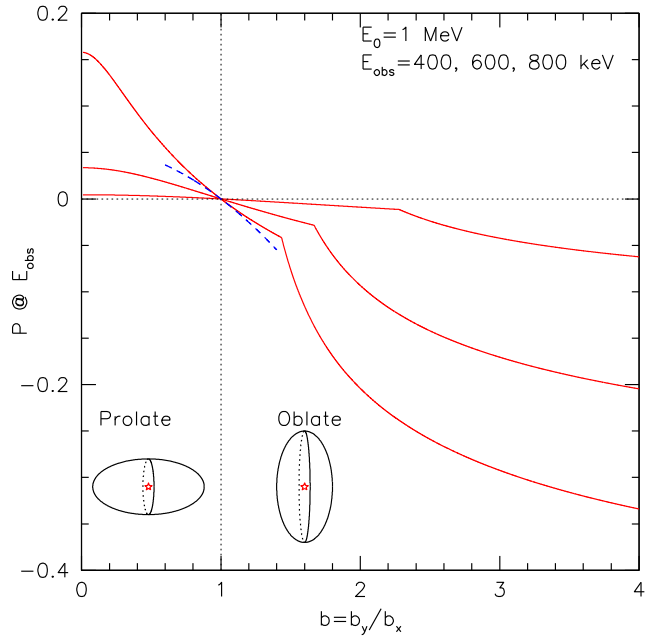}
\end{minipage}
\caption{Polarization of gamma-ray emission scattered on a homogeneous ellipsoid of rotation with a central point source of monochromatic photons with the initial energy $E_0=1$\,MeV, as a function of the ellipsoid's axes ratio $b$. The curves correspond to the observed energies $E_{obs}=400$, 600 and 800 keV (from top to bottom at small $b$). The polarization degree is zero for $b=1$ (sphere) and is positive (negative) for prolate (oblate) ellipsoids (as illustrated by the inset sketches). A kink in the curves is a purely geometrical effect. The dashed blue line corresponds to the lowest order expansion given by eq.~(\ref{eq:blin}).
\label{fig:bpol}
}
\end{figure}

Assuming unpolarized initial radiation,  for a given $\theta$ and $\phi$, the Stokes parameters for the scattered radiation are
\be
I=\frac{1}{2}r_0^2 X^2 \left [X+X^{-1}-\sin^2\theta \right] a(\theta,\phi)
\ee
\be
Q=\frac{1}{2}r_0^2 X^2 \left [\cos 2\phi \sin^2\theta \right] a(\theta,\phi)
\ee
Integration of the above expressions over $\phi$ yields
\be
I=\frac{1}{2} r_0^2 X^2 \left [X+X^{-1}-\sin^2\theta \right] \times 4b E_K\left (-w \right )
\label{eq:bi}
\ee
and
\be
Q=\frac{1}{2} r_0^2 X^2 \times 4b \sin^2\theta\left [ \frac{2E_E(-w)-(2+w)E_K(-w)}{w} \right ]
\label{eq:bq}
\ee
where $w=(b^2-1)\sin^2\theta$, and $E_K$ and $E_E$ are the complete elliptical integrals of the first and the second kind, respectively. 

Values of $I$ and $P$ as a function of $\mu=\cos\theta$ and $E_{obs}$ for the prolate and oblate ellipsoids are shown in Fig.~(\ref{fig:pol_spec}). There, $E_0=1\,{\rm MeV}$ and extreme values of $b=0.1$ and $10$ are used for illustration.  For the prolate ellipsoid [blue curves in Fig.~(\ref{fig:pol_spec})], the 
scattering angles $\sim90$\deg dominate, leading to a peak at $\mu\sim0$ and at $E_{obs}\sim 340$\,keV. These features are also reflected in the degree of polarization. For the oblate ellipsoid, all scattering angles are contributing and the scattered spectrum resembles that of a point source in a sphere. 

The energy dependence of $P$ reflects the intimate relation between the observed energy of the scattered photon and the scattering angle $\theta$, through both the properties of the Compton scattering and the geometry of the problem. At the same time, the spectrum of the reflected continuum directly probes the distribution of scattered photons over $\theta$, while the polarization is also sensitive to the asymmetry in the rotation angle $\phi$, weighted by the $\theta$-dependent coefficient.

For $b=1$, i.e., for a sphere, the polarization degree is zero, of course. For small ellipticity, the lowest order expansion of Eqs.~(\ref{eq:bi}) and (\ref{eq:bq}) in parameter $(b-1)\rightarrow 0$  gives the polarization degree
\begin{eqnarray}
P=-\frac{1}{8}\frac{\sin^4\theta}{\left [ X+X^{-1}-\sin^2\theta\right ]}2\left( b-1\right ).
\label{eq:blin}
\end{eqnarray}
The corresponding function is shown with the blue dashed line in Fig.~(\ref{fig:bpol}) for $E_{obs}=400$~keV. This approximation captures the dependence of the exact solution on energy and $b$ as far as $|b-1|\lesssim 0.2$.

While in Figure~\ref{fig:pol_spec} extreme values of ellipticity are used to emphasize arising polarization signatures, in normal type Ia supernova, much smaller values are expected as indicated by low polarization levels observed in the optical band. For instance,  the models of \citet{2003ApJ...593..788K} for SN~2001el and \citet{2016ApJ...831...79I} for SN~2015bn, have $(b-1)\sim 0.1$. Specializing eq.~(\ref{eq:blin}) for the 847~keV line and 90\deg scattering angle (or, equivalently, for the observed energy $\sim 318$~keV), we get $P\approx 0.13 (b-1)\approx 0.013$, i.e. the continuum polarization at the level of $1\%$. The observed level of polarization will be lower by a factor of few due to the contribution of an unpolarized continuum, e.g., due to 3-photon annihilation continuum (see \S\ref{sec:w7} below). Thus, for the modest levels of ellipticity, the degree of polarization is at the fraction of percent level, similar to the optical band.

\subsection{Homogeneous sphere with an off-center point source}

A similar analysis can be done for a homogeneous scattering sphere, when a point source of 1~MeV photons is shifted by $\Delta$ from the center of the sphere (Fig.~\ref{fig:sphere}). The resulting expressions are given in Appendix~\ref{ap:sphere}. Energy dependence of the polarization degree is the same as for ellipsoid [see eqs.~(\ref{eq:sq}) and (\ref{eq:bq})]. For this geometric setup, the lowest order expansion in  $\Delta$  gives reasonable approximation up to $\Delta\sim0.5-0.6$.

\begin{figure}
\begin{minipage}{0.49\textwidth}
\includegraphics[trim= 1mm 5cm 20mm 2cm,
  width=1\textwidth,clip=t,angle=0.,scale=0.98]{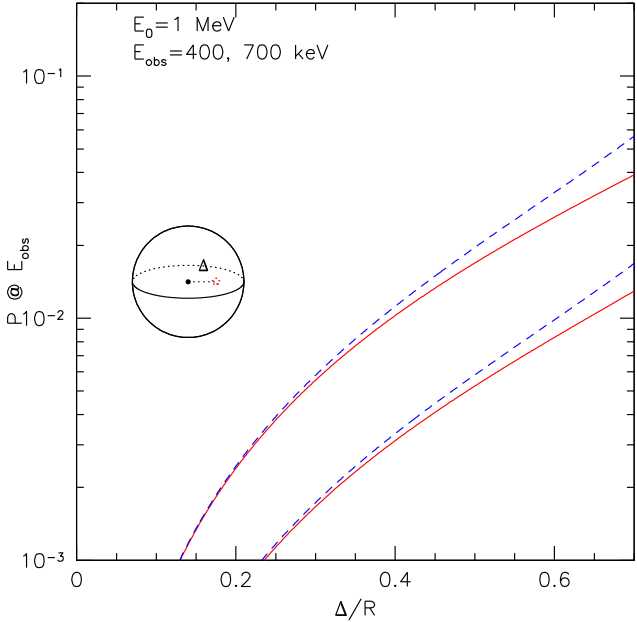}
\end{minipage}
\caption{Polarization of the gamma-ray emission scattered on a homogeneous sphere with an offset point source of monochromatic gamma-ray photons with the initial energy $E_0=1$\,MeV, as a function of the offset $\Delta$. The curves correspond to the observed energies $E_{obs}=400$ and 700 keV (from top to bottom at small $\Delta$).  The dashed blue line shows to the lowest order expansion given by eq.~(\ref{eq:slin}).
\label{fig:sphere}
}
\end{figure}

\begin{figure*}
\begin{minipage}{0.49\textwidth}
\includegraphics[trim= 11cm 1cm 10cm 1cm,
  width=1\textwidth,clip=t,angle=0.,scale=0.98]{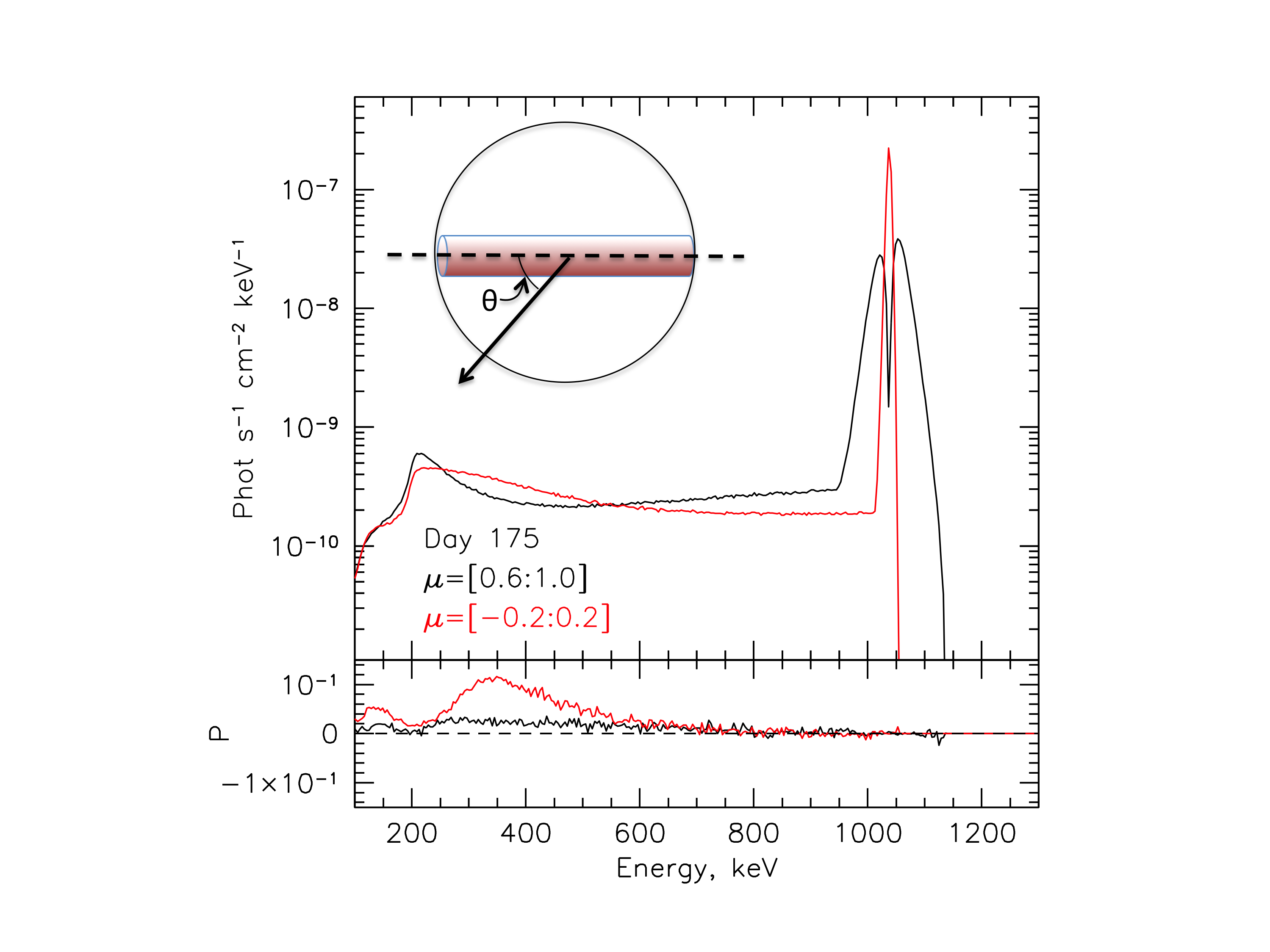}
\end{minipage}
\begin{minipage}{0.49\textwidth}
\includegraphics[trim= 11cm 1cm 10cm 1cm,width=1\textwidth,clip=t,angle=0.,scale=0.98]{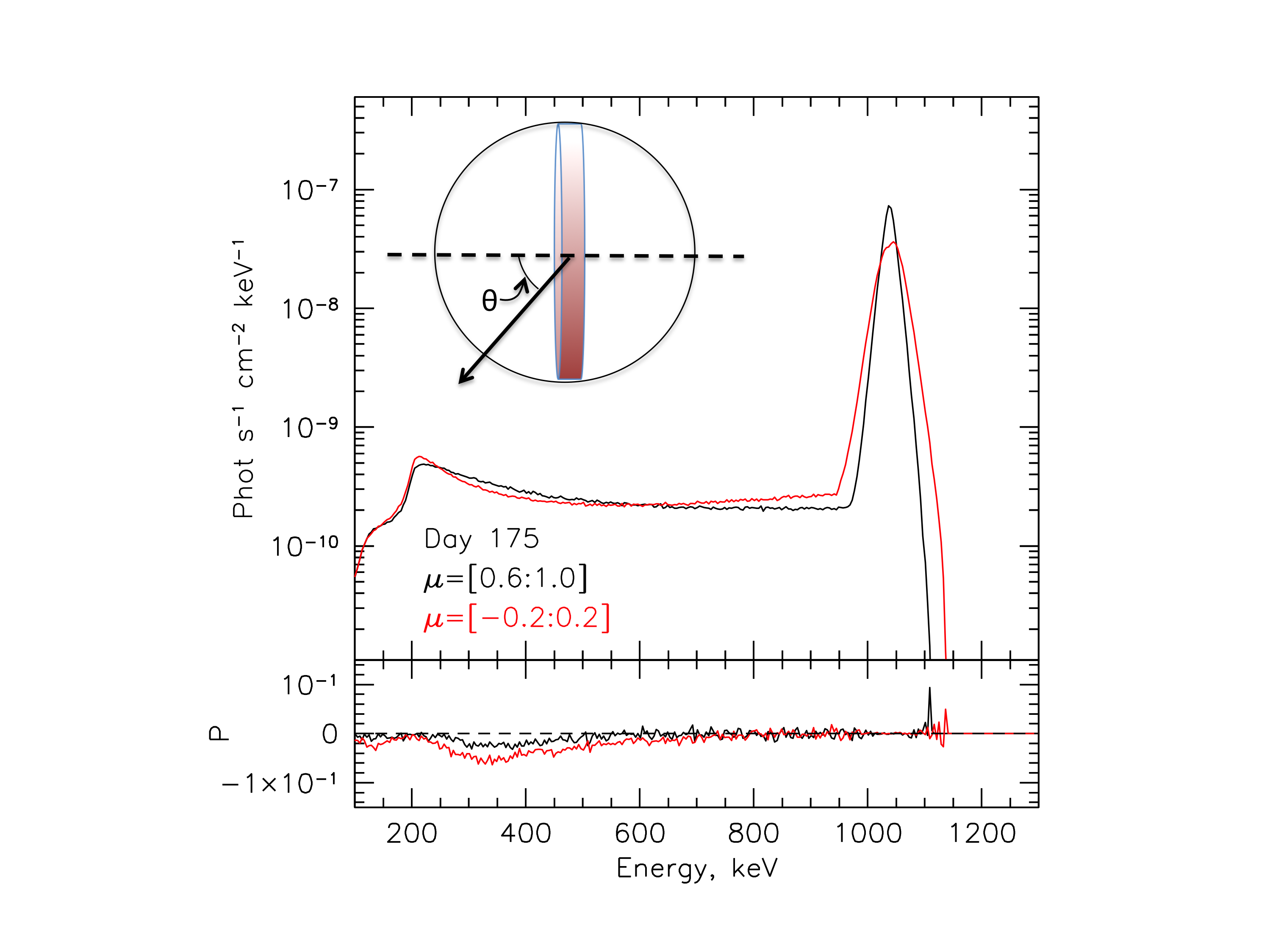}
\end{minipage}
\caption{{\bf Left:} Emerging spectrum (top panel) and polarization degree (bottom panel) for a model, where scattering material has spherically symmetric distribution, while the \co56 is confined to a ``rod''. Only single line at 1238~keV is considered. The opacity effects modify the shape of the line, producing a broad peak with a dip at the core, when viewed at small angle to the rod axis (black curve). When viewing angle is almost perpendicular to the rod, the line becomes narrow (red curve). The legend shows the range of $\mu=\cos\theta$ over which the spectra were averaged in the simulations. Polarization in the continuum (bottom panel) is strongest when the viewing angle is perpendicular to the rod's axis. 
 {\bf Right:} The same as in the left plot, but for \co56 confined to a disk. Both the shape of the line and the degree of polarization are affected. In particular, the degree of polarization is stronger when the viewing angle is perpendicular to the axis of the disc.
\label{fig:pol_line}
}
\end{figure*}

\subsection{Asymmetric distribution of gamma-ray sources in a homologously expanding envelope}
\label{sec:w7}
We now proceed to a slightly more elaborate model, where an asymmetric source of gamma-ray photons is placed in a homologously expanding ejecta. 
The ejecta, consisting of a mixture of Si, S and iron-group elements, have an exponential density profile   
\begin{eqnarray}
\rho(v)=4\,10^2 e^{ \{-v/v_0\}}~{\rm g\,cm^{-3}},
\label{eq:profile}
\end{eqnarray}
where $v_0=3\,10^3~{\rm km~s^{-1}}$. These ejecta represent a spherically symmetric scattering medium with a total mass close to 1.4~$M_\odot$. The radial mass distribution of \co56 is set to be the same as for the scattering medium. The angular distribution is instead
assumed to be highly aspherical, namely in a form of a narrow ``rod'' or a ``disc'', as shown in Fig.\ref{fig:pol_line}. The total mass of \co56 is 0.7~$M_\odot$. Apart from the geometry of the \co56 distribution, the setup is similar to the one used to model gamma-ray emission from SN2014J \citep[see][for details]{2014Natur.512..406C,2015ApJ...812...62C}. 

A Monte Carlo radiative transfer code is used to calculate the emergent spectrum, which includes photoabsorption and Compton scattering (coherent and incoherent).  
As in \S\ref{sec:el} the time delay caused by the propagation of photons through the ejecta is ignored. The simulated spectra are shown in Fig.\ref{fig:pol_line} for the ``rod'' and ``disc'' geometries. The contribution of a single line  at 1037~keV, arising from  \co56 decay, is shown for clarity. As expected, both the spectrum and the degree of polarization show the strong dependence on the viewing angle. The shape of the line reflects the line-of-sight velocity distribution of \co56, which for the rod model changes from a double-peak broad line (when viewing almost along the rod axis) to a single-peak narrow (viewing almost perpendicular to the rod axis). For the disc geometry, the changes in the line shape are also present, but they are less pronounced.

\begin{figure}
\begin{minipage}{0.49\textwidth}
\includegraphics[trim= 1mm 5cm 20mm 2cm,
  width=1\textwidth,clip=t,angle=0.,scale=0.98]{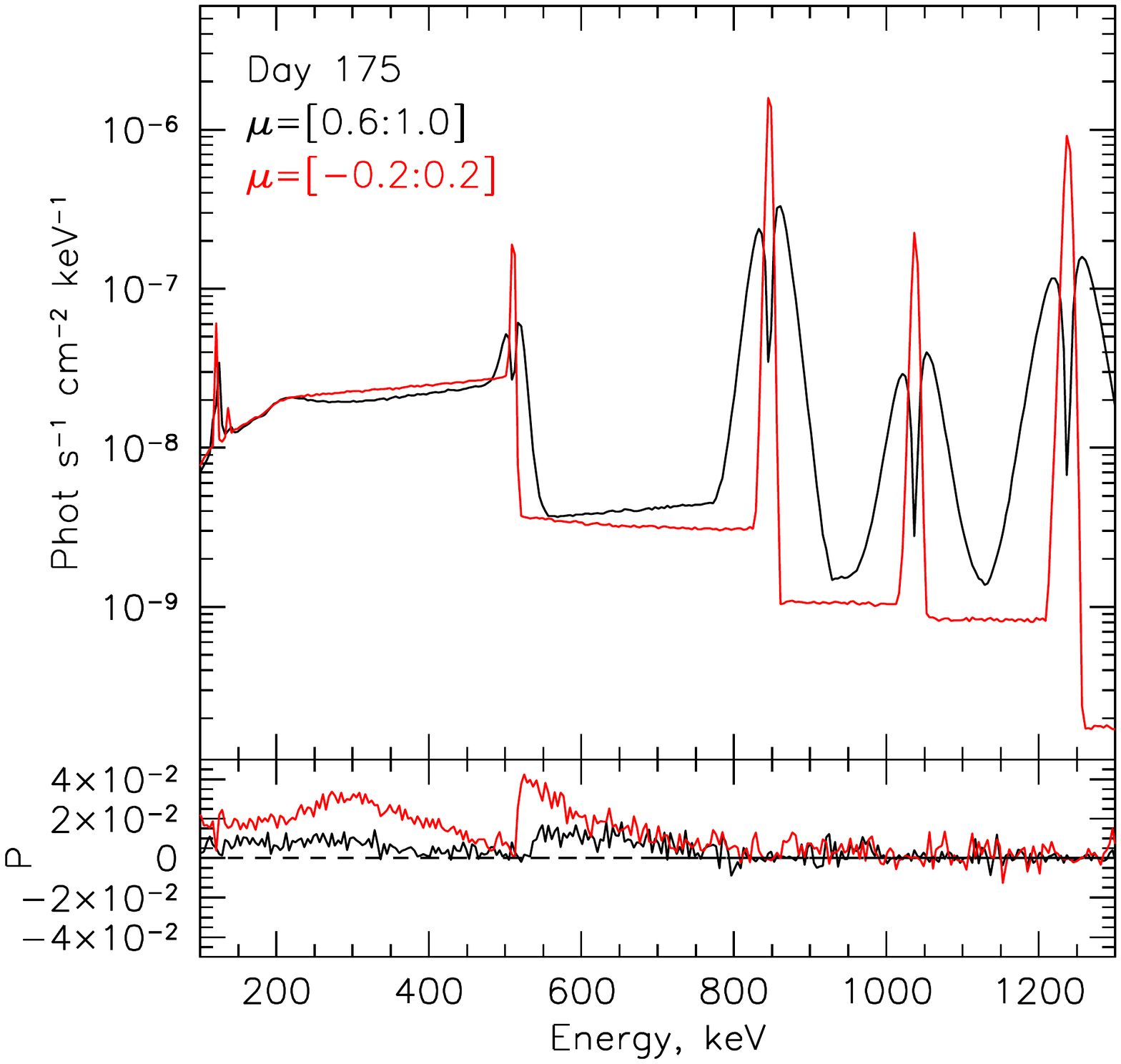}
\end{minipage}
\caption{Same as in Fig.~\ref{fig:pol_line} but with all strong lines and the annihilation emission taken into account. 
\label{fig:pol_lines}
}
\end{figure}

With regard to polarization, it is naturally low when the object is viewed close to the symmetry axis and reaches 
5-10\% when the line-of-sight is perpendicular to it. The  calculation was made for day 175 after the explosion, so that the opacity is relatively low and a single scattering approximation is sufficient\footnote{Although we invoke the single-scattering approximation to interpret the polarization properties, the code does take multiple scatterings into account.} to interpret the energy dependence of the polarization degree (see \S\ref{sec:el} and Figure~\ref{fig:pol_spec}). The spectrum of a singly-scattered continuum has a characteristic ``bowl'' shape with two peaks -- one near the initial energy of the line (corresponding to the forward scattering) and another one close 200~keV,  associated with the backscattering and the largest recoil effect.  The polarization degree is  low for the forward and backward scattering; it is expected to reach maximum for scattering angles $\sim$90\deg, corresponding to energies $\sim$340~keV (for the initial photon energy 1037~keV), as is indeed seen in  Fig.\ref{fig:pol_line}. 

Fig.\ref{fig:pol_lines} shows the spectrum and the degree of polarization for the same ``rod'' model when contributions from all major lines of \co56 and {$^{57}{\rm Co}$} are combined, along with the contribution of the annihilation line at 511~keV and the 3-photon continuum (below 511 keV). The degree of polarization decreases to $\sim$4\% due to the overlap of the scattered continua from different lines. The strongest polarization is at $\sim$300~keV and, also, just above the 511~keV line. Clearly, this polarized continuum is due to the scattered photons of the strongest \co56 line at 847~keV.

\begin{figure}
\begin{minipage}{0.49\textwidth}
\includegraphics[trim= 1mm 5cm 20mm 2cm,
  width=1\textwidth,clip=t,angle=0.,scale=0.98]{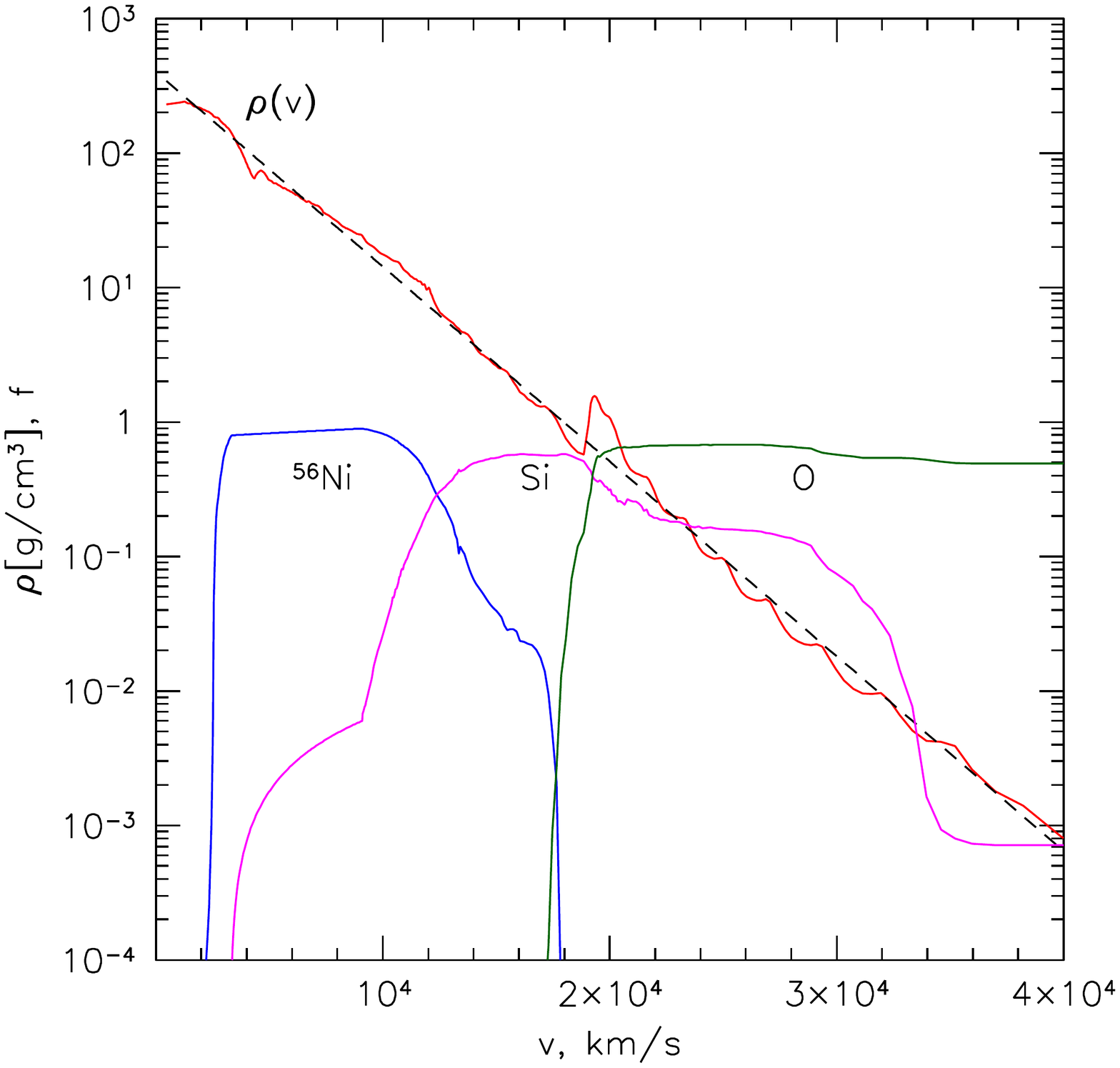}
\end{minipage}
\caption{Density of the ejecta and mass fractions of several major spices as a function of the velocity in the DDC0 model from \citet{2013MNRAS.429.2127B}. The red line is the total density in the model, while the dashed black line is the exponential profile, given by eq.~\ref{eq:profile}. The blue, magenta and green lines show the mass fractions of \ni56, Si and O, respectively.
\label{fig:profile}
}
\end{figure}

\section{Propagation and escape of positrons}
\label{sec:pos}
In the previous section, gamma-ray polarization was principally driven by the asymmetry in the angular distributions of the gamma-ray lines sources and/or of the scattering medium. In this section,   we assume, instead, a perfectly spherically symmetric model and consider how the shape of the 511~keV line is affected by the propagation of positrons through the ejecta. The simplest model that captures the impact of positrons propagation on the line profile can be build as follows.  Let us assume that the positrons are born with initial energy $\sim{\rm MeV}$. They first slow down via ionization losses and, also, via Coulomb collisions with the ejecta electrons \citep[e.g.][]{1979ApJ...228..928B,1993ApJ...405..614C}. During this slow-down process, the positrons can move through the ejecta. Once their energy is in the eV range, they annihilate, producing a 511~keV line that reflects the ejecta velocity distribution over the annihilation site, which in principle might be different from the parent velocity distribution of \co56.

\begin{figure}
\begin{minipage}{0.49\textwidth}
\includegraphics[trim= 1mm 5cm 20mm 2cm,
  width=1\textwidth,clip=t,angle=0.,scale=0.98]{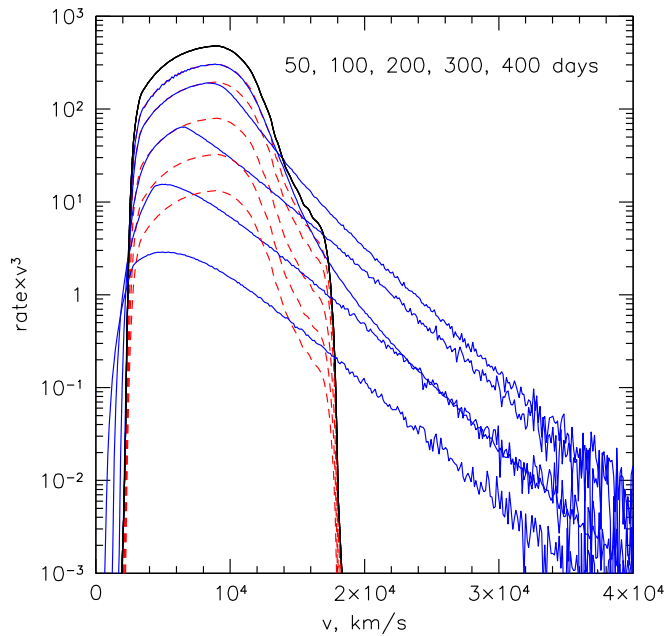}
\end{minipage}
\caption{Distribution of gamma line sources ($\propto \rho_{56}\times v^3$, dashed red lines) and the annihilating positrons (blue lines) for 50, 100, 200, 300, and 400 days since the explosion. The production of positrons  has the same profile as the sources of gamma-ray lines. The subsequent propagation of positrons broadens the velocity distribution of the annihilating positrons. In this toy model, the positrons are moving along the straight lines until they slow down and then annihilate. This limit corresponds to zero magnetic field.
\label{fig:nomag_distr}
}
\end{figure}

The escape of positrons from the SNIa ejecta has been considered in a number of studies \citep[e.g.][]{1993ApJ...405..614C,1998ApJ...500..360R,1999ApJS..124..503M,2014ApJ...795...84P}, which thoroughly consider the process of the positrons slow-down for various topologies of the magnetic field and various distributions of  radioactive isotopes in the ejecta. Here, we consider a spherically-symmetric homologously-expanding ejecta in which \co56 (or initially \ni56) is predominantly found in the inner layers. To this end, we use the publicly available DDC0 model from \citet{2013MNRAS.429.2127B} as an example. The density profiles and the mass fractions of a few major species are shown in Fig.~\ref{fig:profile}. The corresponding distribution of produced gamma-ray photons (and positrons) is shown with the black line in Fig.~\ref{fig:nomag_distr}, assuming that all \ni56 is converted to \co56. In this Figure, the density of \co56 is multiplied by $v^3$ to show the range of velocities making the dominant contribution to the produced photons. With time the amount of \co56 goes down (after day $\sim$25), but the shape of the distribution does not change. Corresponding curves are shown with the red dashed lines for days 50, 100, 200, 300, 400.

\begin{figure}
\begin{minipage}{0.49\textwidth}
\includegraphics[trim= 1mm 5cm 20mm 2cm,
  width=1\textwidth,clip=t,angle=0.,scale=0.98]{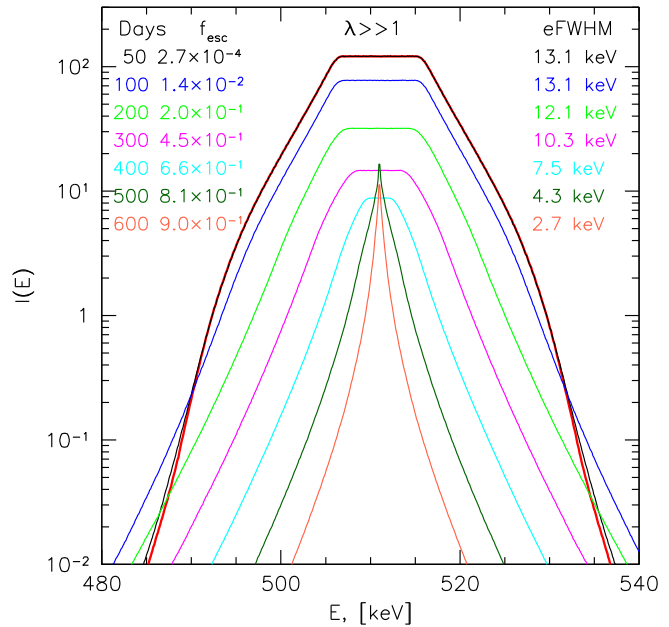}
\end{minipage}
\caption{Velocity distribution of ejecta, where the positrons annihilate for $\lambda\gg1$ for days 50, 100, 200, 300, 400, 500, 600. In this model, positrons propagate along the straight lines. The curves are calculated as the 511 keV line spectrum, assuming that annihilation of each positron leads to a monochromatic 511~keV line in the local ejecta frame. For comparison, the ``local'' model, which assumes that positrons annihilate where they are produced is shown with the thick red line for day 50. As long as the column density of the eject is very large, the migration of positrons over radius is sub-dominant and the core of the line maintains its flat-top structure with the eFWHM$=13$~keV. As the column density becomes comparable to $L_{in}$, the line gradually develops a narrow core and broad wings due to migration and escape of the positrons.
\label{fig:ann_days}
}
\end{figure}

The radial optical depth for gamma-ray photons drops below unity around day 50-100, depending on the particular ejecta model and on the energy of the photons. In the limit of small optical depth, the expected line profiles will reflect the parent isotope velocity distribution projected to the line of sight.  Therefore, all lines should have the same profile, once plotted as a function of $E_{obs}/E_{0}$, where $E_0$ is the initial photon energy. As an example, we show the profile of a line with $E_0=511$~keV for day 50 in  Fig.~\ref{fig:ann_days}. The flat-top shape is caused by the lack of \co56 at small velocities in the ejecta model shown in  Fig.~\ref{fig:profile}.

Positrons are produced in $\sim$19\% of \co56 decays\footnote{We do not consider here the positrons produced by the ${^{44}}{\rm Ti}$ decay, which is important for the much later phases of the supernova evolution} with the kinetic energy between 0 and  1459~keV  \citep[see, e.g,][for the discussion of \co56 decay]{1994ApJS...92..527N}. We assume that all positrons have the same initial energy $E_{in}=0.632$~MeV, which corresponds to the mean energy of the \co56 positrons.  For such initial energy, the annihilation rate is much smaller than the energy loss rate for ionization and Coulomb collisions. Therefore, the positrons will first cool down and then annihilate. From this point of view, the initial energy of the positron has no influence on the energy of the annihilating positron. However, the initial energy $E_{in}$ determines the cool-down time of the positron and, therefore, the distance it can travel from the place where it was born. For trans-relativistic electrons or positrons, the rate of energy losses in the neutral or weakly ionized medium does not depend strongly on the properties of the medium. For estimates, we have adopted an energy losses rate, per unit mass column density, at $E=E_{in}$ as
\be
\frac{d E}{dL}\approx 1.6~{\rm MeV\,g^{-1}\,cm^{2} },
\ee
which corresponds to the energy losses of a 600~keV positron in a neutral Si (NIST Standard Reference Database 124, Berger et al.).  The value $L_{in}=\displaystyle E_{in}/\frac{d E}{d\rho x}\approx 0.4~{\rm g\,cm^{-2}}$ sets the mass column density the positron has to cover before its energy changes by $\Delta E\sim E_{in}$.  Since the losses increase once the positron becomes non-relativistic, the value of $L_{in}$ can be used as an estimate of the  total column density needed for the positron to slow-down to energies $\ll E_{in}$. 

Once $L_{in}$ is known, it is straightforward to calculate the distance a positron will travel before it slows down.
To allow for a qualitative account for a tangled magnetic field in the model, we introduce a parameter $\lambda$, which is the effective mean free path (for the change of direction) of the positron during the slow-down phase. In the model, $\lambda$ is measured in units of the maximum radius of the homologously  expanding ejecta, corresponding to $v_{max}\times t$, where $v_{max}=4\,10^4\,{\rm km\,s^{-1}}$. With this definition, $\lambda \gg 1$ corresponds to positrons propagating along straight lines, set by their initial directions. On the contrary, $\lambda \ll 1$ implies that the positron changes direction after travelling a distance equal to $\lambda$. For the  $\lambda \ll 1$, the adiabatic losses, associated with the expansion of the envelope, will also contribute to the slow-down process. However, for the ejecta evolutionary phases considered here, the adiabatic losses can be neglected.

This simplistic scheme of accounting for the motion of electrons in the magnetic field, has, of course, several obvious shortcomings. 
For instance, the gyration of a positron around field lines would increase the effective column density accumulated by the positron while passing the same length along the field line.
Magnetic mirrors could also play a role by either suppressing the positrons transport (tangled field) or by promoting radial outward transport \citep[for the radially combed field lines, see, e.g.][]{1993ApJ...405..614C} and, simultaneously, by preventing penetration of the positrons into the core. Nevertheless, the $\lambda$ parameter can be used to illustrate the difference between the two limits of i) free propagation of positrons and ii) the positrons locked to the place, where they were produced.

We further assume, that once the positron slows down, it annihilates immediately (see discussion below). Therefore, to find the velocity distribution of the ejecta where positrons are annihilating, it is sufficient to (i) calculate the probability of a positron produced at a given initial radius to accumulate the mass column density $L_{in}$ at given final radius, and (ii) convolve this distribution with the distribution of \co56.  To model the transport of positrons we used a Monte Carlo method, which allowed us to handle both limits of large and small mean-free-path of positrons using the same code. Also, this approach simplified the modeling of the gradual evolution of positrons' energies due to ionization losses.
The migration of positrons along the radius is illustrated in Fig.~(\ref{fig:nomag_distr}). The red dashed lines in this Figure show the distribution of \co56 (red dashed lines) for days 50, 100, 200, 300, 400.  The decrease of the overall normalization is due to the gradual decay of \co56. For this plot, we assumed that  $\lambda \gg 1$, i.e., the positrons move along straight lines.  With this approximation, the only parameter that defines the magnitude of positrons migration is the total mass column density of the ejecta $L_{tot}$ in comparison with $L_{in}$. The total column density of the ejecta (from the center to infinity) in the considered model is $L_{tot}\sim 15~{\rm g\,cm^{-2}}$ at day 100 and it scales as $\displaystyle \left ( \frac{t}{100\,{\rm days}}\right )^{-2}$. The blue lines show the distribution of annihilating positrons. For $t=$50--100 days, $L_{in}$ is much smaller than $L_{tot}$,  the effect of migration is small and the blue lines are similar to the corresponding red lines. For later epochs, the ejecta progressively become more and more ``transparent'' for positrons (especially, the outer parts), so instead of slowing-down and annihilating locally, the positrons migrate to larger and smaller radii. Thus, some positrons will annihilate in the ejecta having much larger or much smaller velocities than those associated with the bulk of \co56.

\begin{figure}
\begin{minipage}{0.49\textwidth}
\includegraphics[trim= 1mm 5cm 20mm 2cm,
  width=1\textwidth,clip=t,angle=0.,scale=0.98]{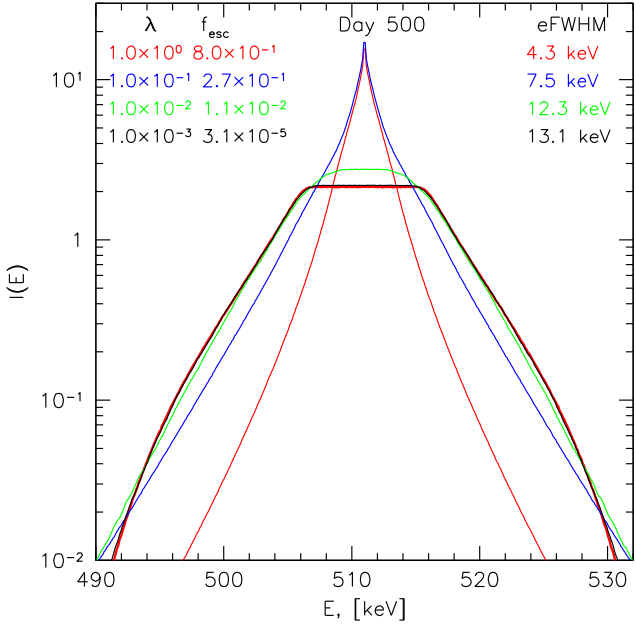}
\end{minipage}
\caption{Impact of the small mean-free-path on the shape of the line for day 500. By that time the radial column density of the ejecta is already comparable with $L_{in}$, which means that the positrons, moving along straight lines can escape the ejecta before annihilation. The role of small $\lambda$ is to replace the straight trajectories with the diffusion and increase the effective column density traversed by the positron. As a result, the effect of radial migration goes down and the line recovers the flat-top shape, characteristic for \co56 distribution. 
\label{fig:mfp_line}
}
\end{figure}


By day 500, $L_{tot}$ becomes comparable to $L_{in}$, i.e. the positrons can cross the ejecta before they slow down. Even earlier, the positrons can escape from the radii, where the bulk of \co56 is located. 

It is convenient to illustrate the velocity distributions shown in Fig.~(\ref{fig:nomag_distr}) by calculating the expected 511 keV line spectra, under the assumption that annihilation of each positron leads to a pair of monochromatic 511~keV photons (in the local reference frame of the ejecta). Corresponding spectra are shown in   Fig.~(\ref{fig:ann_days}) for days 50, 100, 200, 300, 400, 500, and 600. It is clear that once the mass column density of the ejecta becomes small, the line changes from a broad flat-top structure (reflecting \co56 distribution) to a narrow peak. This narrow peak is due to  the positrons, which trajectories happened to cross the densest (low velocity) core of the ejecta. Most of the positrons moving in other directions simply escape from the ejecta, leading to the significant decrease of the line flux and the disappearance of the flat-top part of the line.   Note, however, that, while the core becomes more narrow, the very far wings of the line become more pronounced due to the migration of positrons to large radii.    
The fraction of escaping positrons and the effective FWHM\footnote{The effective Full-Width-Half-Maximum (eFWHM) is defined as an energy interval containing 76\% of photons} of the line are shown in Fig.~(\ref{fig:ann_days}). 

We now consider a case when the mean-free-path of positrons is smaller than the size of the ejecta, i.e. $\lambda\lesssim 1$ for day 500, as shown in Fig.~(\ref{fig:mfp_line}). As is clear from this Figure, the decrease of $\lambda$ forces the line to become broader and broader. For $\lambda=10^{-3}$ the velocity distribution is very close to that of \co56. Such behaviour is easy to understand. Indeed, for small $\lambda$, the trajectory of the positron becomes a 3D random walk with very small steps. This means that even if the radial column density of the ejecta is small (as is the case for day 500), the positron will be able to slow down during long diffusion time. As a result, the positron will annihilate close to the place where it was produced. Of course, for early stages of the ejecta expansion, e.g., day 100, the positrons will accumulate the column density $\sim L_{in}$ and slow down after travelling small  distance in the ejecta. In this case, the value of $\lambda$ does not have any impact on the final velocity distribution. For the late epochs, the random walk increases the effective column density by a factor $\sim \lambda^{-1}$, decreasing the escape fraction and increasing the width of the line.

\begin{figure}
\begin{minipage}{0.49\textwidth}
\includegraphics[trim= 1mm 5cm 20mm 2cm,
  width=1\textwidth,clip=t,angle=0.,scale=0.98]{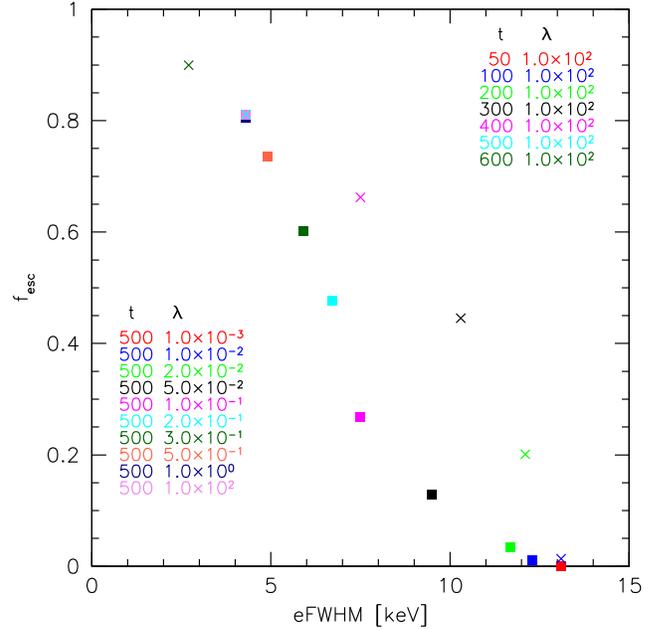}
\end{minipage}
\caption{Relation between the effective line width and the positron escape fraction for a set of models shown in Figures (\ref{fig:ann_days}) and (\ref{fig:mfp_line}). The value of $f_{esc}$ characterizes the fraction of escaping positrons, which are produced at a given time, shown in the legend. Solid squares show the runs for $t=500\,{\rm days}$ and different values of $\lambda$. Crosses shows the runs with the same $\lambda\gg1$, i.e. positrons are propagating along straight lines, but for different moments $t=$50,100,200,300,400,500, and 600 days. As expected, the increase of $\lambda$ and $t$ lead to larger escape fraction and narrower line.
\label{fig:escape}
}
\end{figure}

From Figures (\ref{fig:ann_days}) and (\ref{fig:mfp_line}) it is clear that line width (eFWHM) and the positron escape fraction ($f_{esc}$) are closely related. Indeed, the change of the line shape requires significant radial migration of positrons. This, in turn, implies that a significant fraction of positrons can escape. This is illustrated in Figures (\ref{fig:escape}), which shows the relation between  eFWHM and $f_{esc}$ for the set of simulations shown in Fig.~(\ref{fig:ann_days}) and (\ref{fig:mfp_line}). As eFWHM goes down, the fraction of escaping positrons goes up, although it is not a one-to-one relation when a full range of parameters (i.e., the time since the outburst and the value of $\lambda$) is considered. 

The spectra shown in Figures (\ref{fig:ann_days}) and (\ref{fig:mfp_line}) are calculated assuming that annihilating positrons produce narrow line at 511~keV. In reality, the annihilation process is more complicated, and it always leads to a broadened line and a 3-photon continuum, which is associated with the annihilation via the formation of a bound electron-positron pair (positronium) in a triplet state. The annihilation of positrons in the astrophysical (hydrogen-dominated) plasma has been considered in detail in relation to the annihilation emission from the center of the Milky Way \citep[e.g.][]{1979ApJ...228..928B}. Depending on the ionization fraction and the temperature of the gas, the width and the fraction of the annihilations via positronium formation change in a non-trivial way \citep[see, e.g., Fig.11 in][]{2011MNRAS.411.1727C}. The main difference of the SNIa ejecta, compared to regular astrophysical plasma, is, of course, the domination of heavy elements. Few limiting cases can be qualitatively considered. 

In one limit, the ejecta are assumed to be completely neutral.  For the neutral hydrogen case the expected eFHWM for the positronium formed in-flight is $\sim 5$~eV  \citep[][]{1979ApJ...228..928B}.The ionization potentials $I$ of elements like Si, S or Fe are between 7.9 and 10.4~eV, i.e., lower than in hydrogen.  The threshold in the positron's kinetic energy  for the positronium formation, i.e., $I-6.8\,{\rm eV}$ will be lower too. 
As a result, one can expect the width of the line in the SNIa ejecta (set by the kinetic energy of formed positronium) be smaller than in the cold hydrogen plasma. For estimates, we assume that width of the line is proportional to the mean between the ionization and the positronium formation thresholds, i.e. $\propto (2I-6.8\,{\rm eV})/2$. For $I=8.15\,{\rm eV}$ (corresponding to neutral Si) this scaling predicts the width of the 511~keV line $\sim2.5\,{\rm keV}$. 

In another limit, the annihilation takes place in the gas of free electrons, and the width of the line is a function of temperature only: $\displaystyle \sim 1.1 \left ( T/{10^4\,{\rm K}}\right )^{1/2}\,{\rm keV}$ \citep{1976ApJ...210..582C}. This value is also small compared to the characteristic width of the line spectrum shown in Figures (\ref{fig:ann_days}) and (\ref{fig:mfp_line}). Thus, in both limits, the impact of the ``extra'' width of the 511 keV (caused by the positronium motion relative the local ejecta frame) on the observed spectra is not going to be dramatic. 

The above discussions involves a number of (explicit and implicit) simplifications. For example, we completely ignore (i) the opacity effects for gamma-ray photons; (ii) the time delay, caused by propagation of photons from different parts of the ejecta and (iii) aberration effects due to fast-moving ejecta layers. For the positrons we also assumed that their propagation through the ejecta is essentially instantaneous. All these assumption can be lifted in a slightly more elaborate model, but to the model considered in \S\ref{sec:pol} and \S\ref{sec:pos} suffice to illustrate the main effects.

\section{Conclusions}
\label{sec:conclusions}
We have shown that 
\begin{itemize}
\item Asymmetries in the distribution of the radioactive \co56 , or in the distribution of non-radioactive ejecta, which scatter gamma-ray photons, should lead to an energy-dependent polarization in the scattered continuum.  Such information allows for a non-trivial diagnostic of the ejecta.
\item Propagation of positrons through the ejecta leads to a different shape of the 511 keV line if the positrons annihilate far from the place where they are produced. One could probe the efficiency of the positrons migration through the ejecta by comparing the shape of the annihilation line with other gamma-ray lines, which directly reflect the velocity distribution of \co56. In the setup considered here, the 511 keV line develops a narrow core and broad wings, if positrons are able to move efficiently through the ejecta. A particular shape of the line depends on the distribution of \co56 and on the topology of the magnetic field.
\end{itemize}
The diagnostic discussed here is complementary to the other ways of constraining the properties of the ejecta, e.g., from the light curves and spectra in the optical and gamma-ray bands. For instance, the asymmetries in the line-of-sight distribution of \co56 do not generate polarization, but reveal themselves in the gamma-ray line profiles. On the contrary, the asymmetries in the sky plane are causing polarization in the continuum, but do not generate distortions in the line profiles. Of course, the detection of polarization and the spectrum of 511 keV some 100-500 days after a Type Ia supernova explosion is a more difficult problem than finding and characterizing the brightest gamma-ray lines at the peak of their luminosity (50-100 days after the explosion).  This problem would be lifted for a truly nearby supernova, e.g. in the Milky Way galaxy, at a distance of 10\,kpc. Compared to SN2014J in M82 (at a distance of $\sim 3.5\,{\rm Mpc}$), the local supernova would be some $10^5$ times brighter,  opening a way for much more rich diagnostics. As mentioned in \S\ref{sec:el}, one can expect polarization degree at level below one percent at energies $\sim$300~keV
in the case of modest ellipticity $(b-1) \sim 0.1$. The proposed \ASTROGAM mission at energies $\sim$0.2-2 MeV can achieve a Minimum Detectable Polarization (MDP) at the 99\% confidence level as low as 0.7\% for a Crab-like source in 1~Ms exposure \citep[see][]{2017arXiv171101265D}. The flux density of a nearby supernova at 300 keV would correspond to $\sim$40 Crabs (175 days after the explosion). Therefore, sub-percent levels of the polarization will be within reach. For extreme values of asymmetry, the polarization could be as high as few percents (see, e.g., Fig.~\ref{fig:pol_lines}. As regards the line shapes (see \S\ref{sec:pos}), accurate measurements of the width for all major lines from such a bright source should not be a problem using the currently operating \INTEGRAL mission \citep{2003A&A...411L...1W}.

\section{Acknowledgements}
We are grateful to  N.N.Chugai and L.A.Vainstein for useful discussions.
We acknowledge partial support by grant No. 14-22-00271 from the Russian Scientific Foundation.

\appendix

\section{Polarization for sphere and an offset point source}
\label{ap:sphere}
Here we consider a homogeneous sphere with a point source of gamma-ray photons, which is shifted from the center of the sphere (along the $x$ axis) by $\Delta$, where $\Delta$ is expressed in units of the sphere radius. 
Similar to the ellipsoid case, the length of the vector from the source to the surface of the sphere is
\be
a(\theta,\phi)=-\Delta\sin\theta\cos\phi+\sqrt{1-\Delta^2+\Delta^2\sin^2\theta\cos^2\phi}.
\ee 

Integration of the Klein-Nishina cross section over all $\phi$ yields 
\begin{eqnarray}
I=2 r_0^2 X^2 \left (X+X^{-1}-\sin^2\theta \right) \times \sqrt{1-\Delta^2} E_E(-w)
\end{eqnarray}
and
\begin{eqnarray}
Q=2 r_0^2 X^2 \left [ \frac{(w+2)E_E(-w)-2(w+1)E_K(-w)}{3w}\right ]\sin^2\theta
\label{eq:sq}
\end{eqnarray}
where $\displaystyle w=\frac{\Delta^2}{1-\Delta^2}\sin^2\theta$.

The lowest order expansion in the parameter $\Delta\rightarrow 0$ of eqs.~(\ref{eq:bi}) and (\ref{eq:bq}), gives the polarization degree
\begin{eqnarray}
P=\frac{1}{8}\frac{\sin^4\theta}{\left [ X+X^{-1}-\sin^2\theta\right ]} \frac{\Delta^2}{1-\Delta^2}.
\label{eq:slin}
\end{eqnarray}

\label{ap:model}

\label{lastpage}
\end{document}